\documentclass[technote]{IEEEtran}

\newtheorem{proposition}{Proposition}
\newtheorem{corollary}{Corollary}
\newtheorem{definition}{Definition}
\newtheorem{lemma}{Lemma}

\newcommand{\df}{\stackrel{\mbox{\scriptsize def}}{=}}
\newcommand{\ELS}{ELS}
\newcommand{\rk}{\mathrm{rk}}

\newcommand{\dr}{d_{\mbox{\tiny{R}}}}

\usepackage{amsmath, amssymb,cite}
\usepackage[dvips]{graphicx}

\begin{document}
\title{On the Decoder Error Probability of Bounded Rank-Distance Decoders for Maximum Rank Distance
Codes}

\author{Maximilien Gadouleau,~\IEEEmembership{Student Member, IEEE,}
and Zhiyuan Yan,~\IEEEmembership{Member, IEEE}%
\thanks{The authors are with the Department of Electrical and Computer Engineering, Lehigh University, Bethlehem, PA, 18015 USA (e-mail: magc@lehigh.edu; yan@lehigh.edu).}}

\maketitle

\thispagestyle{empty}

\begin{abstract}
In this paper, we first introduce the concept of \emph{elementary
linear subspace}, which has similar properties to those of a set of
coordinates. We then use elementary linear subspaces to derive
properties of maximum rank distance (MRD) codes that parallel those
of maximum distance separable codes. Using these properties, we show
that, for MRD codes with error correction capability $t$, the
decoder error probability of bounded rank distance decoders
\emph{decreases exponentially} with $t^2$ based on the assumption
that all errors with the same rank are equally likely.
\end{abstract}


\section{Introduction}\label{sec:intro}

Although the rank of a matrix has long been known to be a metric
\cite{hua_cms51}, the rank metric was first considered for error
control codes (ECCs) by Delsarte \cite{delsarte_jct78}. ECCs with
the rank metric \cite{gabidulin_pit0185, roth_it91,
gabidulin_lncs91, lusina_it03} have been receiving growing attention
due to their applications in storage systems \cite{roth_it91},
public-key cryptosystems \cite{gabidulin_lncs91}, space-time coding
\cite{lusina_it03}, and network coding \cite{koetter_2007,
silva_07}.

The pioneering works in \cite{delsarte_jct78, gabidulin_pit0185,
roth_it91} have established many important properties of rank metric
codes. Independently in \cite{delsarte_jct78, gabidulin_pit0185,
roth_it91}, a Singleton bound (up to some variations) on the minimum
rank distance of codes was established, and a class of codes that
achieve the bound with equality was constructed. We refer to
\emph{linear or nonlinear} codes that attain the Singleton bound as
maximum rank distance (MRD) codes, and the class of linear MRD codes
proposed in \cite{gabidulin_pit0185} as Gabidulin codes henceforth.
Different decoding algorithms for Gabidulin codes were proposed in
\cite{gabidulin_pit0185, roth_it91, richter_isit04, loidreau_wcc05}.

In this paper, we investigate the error performance of \emph{bounded
rank distance decoder} for MRD codes. A bounded rank distance
decoder for MRD codes with error correction capability $t$ is
guaranteed to correct all errors with rank no more than $t$. Given a
received word, a bounded rank distance decoder either provides an
estimate for the transmitted codeword or declares decoder failure. A
decoder error occurs when the estimate is not the actual transmitted
codeword. The main results of this paper are new upper bounds on the
decoder error probability (DEP) of bounded rank distance decoders
for MRD codes. We emphasize that the DEP considered herein is
\emph{conditional}: it is the probability that a bounded rank
distance decoder, correcting up to $t$ rank errors, makes an
erroneous correction, given that an error with a fixed rank was
made. Our bounds indicate that the DEP of MRD codes with error
correction capability $t$ \emph{decreases exponentially} with $t^2$.
To derive our bounds, we assume all errors with the same rank are
equally likely.

We provide the following remarks on our results:
\begin{enumerate}
\item Since decoder failures can
be remedied by error masking or retransmission, decoder errors are
more detrimental to the overall performance and hence often
considered separately (see \cite{mceliece_it86}). This is the main
reason we focus on DEP.

\item Note that
bounded rank distance decoders guarantee to correct errors with rank
up to $t$. In \cite{loidreau_acct06}, it was shown that with
Gabidulin codes errors with rank beyond $t$ can be corrected when
errors occur from the same vector space. However, we do not consider
the decoders in \cite{loidreau_acct06} and focus on bounded rank
distance decoders instead.

\item Our bounds are analogous to
the upper bounds on the error probability of bounded Hamming
distance decoders for maximum distance separable (MDS) codes in
\cite{mceliece_it86} (see \cite{driessen_82, cheung_it89,
tolhuizen_eurocode92} for related results).
\end{enumerate}

We are able to derive our bounds based on an approach which
parallels the one in \cite{mceliece_it86}. This was made possible by
the concept of \emph{elementary linear subspace} (ELS), which has
similar properties to those of a set of coordinates. Using
elementary linear subspaces, we also derive useful properties of MRD
codes which parallel those of MDS codes. Although our results may be
derived without the concept of ELS, we have adopted it in this paper
since it enables readers to easily relate our approach and results
to their counterparts for Hamming metric codes.

The rest of the paper is organized as follows.
Section~\ref{sec:preliminaries} gives a brief review of the rank
metric, Singleton bound, and MRD codes. In
Section~\ref{sec:combinatorial}, we derive some combinatorial
properties which are used in the derivation of our upper bounds. In
Section~\ref{sec:lemmas}, we first introduce the concept of
elementary linear subspace and study its properties, and then obtain
some important properties of MRD codes. In
Section~\ref{sec:main_results}, we derive our upper bounds on the
DEP of MRD codes.

\section{Preliminaries}\label{sec:preliminaries}
Consider an $n$-dimensional vector ${\bf x} = (x_0, x_1,\ldots,
x_{n-1}) \in \mathrm{GF}(q^m)^n$. Assume $\{\alpha_0, \alpha_1,
\ldots, \alpha_{m-1}\}$ is a basis of GF$(q^m)$ over GF$(q)$, then
for $j=0, 1, \ldots, n-1$, $x_j$ can be expanded to an
$m$-dimensional column vector $(x_{0, j}, x_{1, j},\ldots, x_{m-1,
j})^T$ over $\mathrm{GF}(q)$ with respect to the basis $\{\alpha_0,
\alpha_1, \ldots, \alpha_{m-1}\}$. Let ${\bf X}$ be the $m\times n$
matrix obtained by expanding all the coordinates of ${\bf x}$. That
is, ${\bf X} = \{x_{i,j}\}_{i,j=0}^{m-1,n-1}$ where $x_j =
\sum_{i=0}^{m-1} x_{i, j}\alpha_i$.  The \emph{rank norm} of the
vector ${\bf x}$ (over GF$(q)$), denoted as $\rk({\bf x})$, is
defined as $\rk({\bf x}) \df \mathrm{rank}({\bf X})$
\cite{gabidulin_pit0185}. The rank norm of ${\bf x}$ is also the
{\em maximum} number of coordinates in ${\bf x}$ that are linearly
independent over GF$(q)$. The field $\mathrm{GF}(q^m)$ may be viewed
as a vector space over $\mathrm{GF}(q)$. The coordinates of ${\bf
x}$ thus span a linear subspace of $\mathrm{GF}(q^m)$, denoted as
$\mathfrak{S}({\bf x})$, such that $\mathrm{dim}\left(
\mathfrak{S}({\bf x}) \right) = \rk({\bf x})$. For all ${\bf x},
{\bf y}\in \mathrm{GF}(q^m)^n$, it is easily verified that $d({\bf
x},{\bf y})\df \rk({\bf x} - {\bf y})$ is a metric over GF$(q^m)^n$,
referred to as the \emph{rank metric} henceforth
\cite{gabidulin_pit0185}. Hence, the {\em minimum rank distance}
$d_{\mbox{\tiny R}}$ of a code is simply the minimum rank distance
over all possible pairs of distinct codewords. A code with a minimum
rank distance $d_{\mbox{\tiny R}}$ can correct all errors with rank
up to $t = \left\lfloor (d_{\mbox{\tiny R}}-1)/2 \right\rfloor$.

The minimum rank distance $d_{\mbox{\tiny R}}$ of a code of length
$n$ over $\mathrm{GF}(q^m)$ satisfies $d_{\mbox{\tiny R}} \leq
d_{\mbox{\tiny H}}$ \cite{gabidulin_pit0185}, where $d_{\mbox{\tiny
H}}$ is the minimum Hamming distance of the same code. Due to the
Singleton bound on the minimum Hamming distance of block codes
\cite{macwilliams_book77}, the minimum rank distance of a block code
of length $n$ and cardinality $M$ over $\mathrm{GF}(q^m)$ thus
satisfies $d_{\mbox{\tiny R}} \leq n-\log_{q^m}M+1.$
In this paper, we refer to this bound as the Singleton bound for
rank metric codes, and to codes that attain the equality as MRD
codes. Note that although an MRD code is not necessarily linear,
this bound implies that its cardinality is a power of $q^m$.

The number of vectors of rank $0 \leq u \leq \min\{m,n\}$ in
$\mathrm{GF}(q^m)^n$ is given by $N_u = {n \brack u} A(m,u)$, where
$A(m,u)$ is defined as follows: $A(m,0) = 1$ and $A(m,u) =
\prod_{i=0}^{u-1}(q^m-q^i)$ for $u \geq 1$. The ${n \brack u}$ term
is the Gaussian binomial~\cite{andrews_book76}, defined as ${n
\brack u} = A(n,u)/A(u,u)$. Note that ${n \brack u}$ is the number
of $u$-dimensional linear subspaces of $\mathrm{GF}(q)^n$
\cite{andrews_book76}.

Note that following the approach in \cite{gabidulin_pit0185}, the
vector form over $\mathrm{GF}(q^m)$ is used to represent rank metric
codes although their rank weight is defined by their corresponding
$m \times n$ code matrices over $\mathrm{GF}(q)$. Naturally, rank
metric codes can be studied in the matrix form (see
\cite{delsarte_jct78, roth_it91}). The vector form is chosen in this
paper since our results and their derivations for rank metric codes
can be related to their counterparts for Hamming metric codes.

\section{Combinatorial Results}\label{sec:combinatorial}
In this section, we derive some combinatorial properties which will
be instrumental in the derivation of our results in
Section~\ref{sec:main_results}.

\begin{lemma}\label{lemma:bound_Amu}
For $0 \leq u \leq m$, $K_q q^{mu} < A(m,u) \leq q^{mu}$, where $K_q
= \prod_{j=1}^{\infty} (1-q^{-j})$. Also, for $0 \leq u \leq m-1$,
$A(m,u) > \frac{q}{q-1} K_q q^{mu}$. Finally, for $0 \leq u \leq
\lfloor m/2 \rfloor$, $A(m,u) \geq \frac{q^2-1}{q^2}q^{mu}$.
\end{lemma}

\begin{proof} The upper bound is trivial. We now prove the lower bounds.
We have $A(m,u) = q^{mu} \prod_{i=0}^{u-1} (1-q^{i-m}) > q^{mu}
K_q$. The second lower bound follows from $A(m,m-1) = \frac{q}{q-1}
q^{-m}A(m,m)$.

The third lower bound clearly holds for $m=0$ and $m=1$. Let us
assume $m \geq 2$ henceforth and denote $\frac{q^{mu}}{A(m,u)}$ as
$D(m,u)$. It can be easily verified that $D(m,u)$ is an increasing
function of $u$. Thus, it suffices to show that $D(m,\lfloor m/2
\rfloor) \leq \frac{q^2}{q^2-1}$ for $m \geq 2$. First, if $m$ is
odd, $m=2p+1$, it can be easily shown that $D(2p+1,p) < D(2p,p)$.
Hence, we need to consider only the case where $m=2p$, with $p \geq
1$. Let us further show that $D(2p,p)$ is a monotonically decreasing
function of $p$ since
\begin{flalign*}
    &D(2p+2,p+1)\\
    &=\frac{q^{2p+2}}{q^{2p+2}-1}\cdot
    \frac{q^{2p+2}}{q^{2p+2}-q} \cdot
    \frac{q^{2p+2}-q^{p+1}}{q^{2p+2}} D(2p,p)\\
    &=\frac{q^{2p+1}(q^{2p+2}-q^{p+1})}{(q^{2p+1}-1)(q^{2p+2}-1)}
    D(2p,p).
\end{flalign*}

The maximum of $D(2p,p)$ is hence given by $D(2,1) =
\frac{q^2}{q^2-1}$.
\end{proof}
It is worth noting that $K_q$ above represents the fraction of
invertible $m\times m$ matrices over $\mathrm{GF}(q)$ as $m$
approaches infinity, and that $K_q$ increases with $q$.

\begin{corollary}\label{cor:bound_gaussian}
For $0 \leq t \leq n$, we have ${n \brack t} < K_q^{-1} q^{t(n-t)}$.
\end{corollary}
\begin{proof}
By definition, ${n \brack t} = A(n,t)/A(t,t)$. Since $A(n,t) \leq
q^{nt}$ and by Lemma~\ref{lemma:bound_Amu}, $A(t,t) > K_q q^{t^2}$,
we obtain ${n \brack t} < K_q^{-1} q^{t(n-t)}$.
\end{proof}

\section{Properties of MRD codes}\label{sec:lemmas}
Many properties of MDS codes are established by studying sets of
coordinates. These sets of coordinates may be viewed as linear
subspaces which have a basis of vectors with Hamming weight $1$.
Similarly, some properties of MRD codes may be established using
elementary linear subspaces (ELS's), which can be considered as the
counterparts of sets of coordinates.

\subsection{Elementary linear subspaces}\label{sec:ELS}
It is a well-known fact in linear algebra (see, for example,
\cite{gabidulin_pit0185}) that a vector ${\bf x}$ of rank $\rk({\bf
x}) \leq u$ can be represented as ${\bf x} =
(x_0,x_1,\ldots,x_{n-1}) = (e_0, e_1,\ldots, e_{u-1}) {\bf A}$,
where $e_j \in \mbox{GF}(q^m)$ for $j=0,1,\ldots,u-1$ and ${\bf A}$
is a $u \times n$ matrix over $\mathrm{GF}(q)$ of full rank $u$. The
concept of elementary linear subspace can be introduced as a
consequence of this representation. However, due to its usefulness
in our approach we define the concept formally and study its
properties below  from a different perspective.

\begin{definition}[Elementary linear subspace]\label{def:ELS}
A linear subspace $\mathcal{V}$ of $\mathrm{GF}(q^m)^n$ is said to
be elementary if it has a basis $B$ consisting of row vectors in
$\mathrm{GF}(q)^n$. $B$ is called an elementary basis of
$\mathcal{V}$. For $0 \leq v \leq n$, we define $E_v(q^m,n)$ as the
set of all \ELS{}'s with dimension $v$ in $\mathrm{GF}(q^m)^n$.
\end{definition}

By definition, a linear subspace $\mathcal{V}$ with dimension $v$ is
an ELS if and only if it is the row span of a $v \times n$ matrix
${\bf B}$ over $\mathrm{GF}(q)$ with full rank. Thus there exists a
bijection between $E_v(q^m,n)$ and $E_v(q,n)$, and $|E_v(q^m,n)| =
{n \brack v}$. Also, it can be easily shown that a linear subspace
$\mathcal{V}$ of $\mathrm{GF}(q^m)^n$ is an ELS if and only if there
exists a basis consisting of vectors of rank $1$ for $\mathcal{V}$.

Next, we show that the properties of ELS's are similar to those of
sets of coordinates.

\begin{proposition}\label{prop:complementary_ELS}
For all $\mathcal{V} \in E_v(q^m,n)$ there exists $\bar{\mathcal{V}}
\in E_{n-v}(q^m,n)$ such that $\mathcal{V} \oplus \bar{\mathcal{V}}
= \mathrm{GF}(q^m)^n$, where $\mathcal{V} \oplus \bar{\mathcal{V}}$
denotes the direct sum of $\mathcal{V}$ and $\bar{\mathcal{V}}$.
\end{proposition}

\begin{proof}
Clearly, the ELS $\bar{\mathcal V}$ having elementary basis
$\bar{B}$ such that $B \cup \bar{B}$ is a basis of
$\mathrm{GF}(q)^n$ satisfies $\mathcal{V} \oplus \bar{\mathcal V} =
\mathrm{GF}(q^m)^n$.
\end{proof}

We say that $\bar{\mathcal{V}}$ is an {\em elementary complement} of
$\mathcal{V}$. Even though an elementary complement always exists,
we remark that it may not be unique.

The diameter of a code for the Hamming metric is defined in
\cite{macwilliams_book77} as the maximum Hamming distance between
two codewords. Similarly, we can define the rank diameter of a
linear subspace.
\begin{definition}\label{def:rank_ls}
The rank diameter of a linear subspace $\mathcal{L}$ of
$\mathrm{GF}(q^m)^n$ is defined to be the maximum rank among the
vectors in $\mathcal{L}$, i.e., $\delta(\mathcal{L}) \df \max_{{\bf
x} \in \mathcal{L}}\{\rk({\bf x})\}.$
\end{definition}

\begin{proposition}\label{prop:rk=dim}
For all $\mathcal{V} \in E_v(q^m,n)$, $\delta(\mathcal{V}) \leq v$.
Furthermore, if $v \leq m$, then $\delta(\mathcal{V}) = v$.
\end{proposition}
\begin{proof}
Any vector ${\bf x} \in \mathcal{V}$ can be expressed as the sum of
at most $v$ vectors of rank $1$, hence its rank is upper bounded by
$v$. Thus, $\delta(\mathcal{V}) \leq v$ by
Definition~\ref{def:rank_ls}. If $v\leq m$, we show that there
exists a vector in $\mathcal{V}$ with rank $v$. Let $B = \{ {\bf
b}_i \}_{i=0}^{v-1}$ be an elementary basis of $\mathcal{V}$, and
consider ${\bf y} = \sum_{i=0}^{v-1} \alpha_i {\bf b}_i$, where
$\{\alpha_i \}_{i=0}^{m-1}$ is a basis of $\mathrm{GF}(q^m)$ over
GF$(q)$. If we expand the coordinates of ${\bf y}$ with respect to
the basis $\{ \alpha_i \}_{i=0}^{m-1}$, we obtain ${\bf Y} =
\left({\bf b}_0^T,\ldots,{\bf b}_{v-1}^T,{\bf 0}^T,\ldots, {\bf
0}^T\right)^T.$ Since the row vectors ${\bf b}_0, {\bf b}_1, \cdots,
{\bf b}_{v-1}$ are linearly independent over $\mathrm{GF}(q)$, ${\bf
Y}$ has rank $v$ and $\rk({\bf y}) = v$.
\end{proof}

\begin{lemma}\label{lemma:x_in_A}
A vector ${\bf x} \in \mathrm{GF}(q^m)^n$ has rank $\leq u$ if and
only if it belongs to some $\mathcal{A} \in E_u(q^m,n)$.
\end{lemma}
\begin{proof}
The necessity is obvious. We now prove the sufficiency. Suppose
${\bf x} = (x_0, x_1,\ldots, x_{n-1})$ has rank $u$, and without
loss of generality, assume its first $u$ coordinates are linearly
independent. Thus, for $j=0, \cdots, n-1$,
$x_j=\sum_{i=0}^{u-1}a_{ij}x_i$, where $a_{ij}\in \mathrm{GF}(q)$.
That is, ${\bf x}=(x_0, x_1,\ldots, x_{u-1}){\bf A}$, where ${\bf
A}=\{a_{ij}\}_{i=0, j=0}^{u-1, n-1} = ({\bf a}_0^T,\ldots,{\bf
a}_{u-1}^T)^T$. Thus ${\bf x} = \sum_{i=0}^{u-1} x_i{\bf a}_i$, with
${\bf a}_i \in \mathrm{GF}(q)^n$ for $0 \leq i \leq u-1$. Let
$\mathcal{A}$ be the ELS of $\mathrm{GF}(q^m)^n$ spanned by ${\bf
a}_i$'s, then $\mathrm{dim}(\mathcal{A})=u$ and ${\bf x} \in
\mathcal{A}$. This proof can be easily adapted to the case where
$\rk({\bf x}) < u$.
\end{proof}

Let $\mathcal{L}$ be a linear subspace of $\mathrm{GF}(q^m)^n$ and
let $\bar{\mathcal L}$ be complementary to $\mathcal{L}$, i.e.,
$\mathcal{L} \oplus \bar{\mathcal{L}} = \mathrm{GF}(q^m)^n$. We
denote the projection of ${\bf x}$ on $\mathcal{L}$ along
$\bar{\mathcal{L}}$ as ${\bf x}_\mathcal{L}$ \cite{roman_book05}.
Remark that ${\bf x} = {\bf x}_\mathcal{L} + {\bf
x}_{\bar{\mathcal{L}}}$. Note that for any given linear subspace
$\mathcal{L}$, its complementary linear subspace $\bar{\mathcal{L}}$
is not unique. Thus, ${\bf x}_\mathcal{L} $ depends on both
$\mathcal{L}$ and $\bar{\mathcal{L}}$, and is well-defined only when
both $\mathcal{L}$ and $\bar{\mathcal{L}}$ are given. All the
projections in this paper are with respect to a pair of fixed linear
subspaces complementary to each other.

\begin{definition}\label{def:vanish}
Let ${\bf x} \in \mathrm{GF}(q^m)^n$ and $\mathcal{L}$ be a linear
subspace. The vector ${\bf x}$ {\em vanishes} on $\mathcal{L}$ if
there exists a linear subspace $\bar{\mathcal{L}}$ complementary to
$\mathcal{L}$ such that ${\bf x} = {\bf x}_{\bar{\mathcal{L}}}$.
\end{definition}

\begin{lemma}\label{lemma:vanish}
A vector ${\bf x} \in \mathrm{GF}(q^m)^n$ has rank $\leq u$ if and
only if it vanishes on some $\mathcal{B} \in E_{n-u}(q^m,n)$.
\end{lemma}
\begin{proof}
Suppose ${\bf x}$ has rank $\leq u$. By Lemma~\ref{lemma:x_in_A},
there exists $\mathcal{A} \in E_u(q^m,n)$ such that ${\bf x} \in
\mathcal{A}$. Let $\bar{\mathcal{A}}$ be an elementary complement of
$\mathcal{A}$. Thus, ${\bf x}$ vanishes on $\bar{\mathcal{A}}$ by
definition. Also, suppose ${\bf x}$ vanishes on an \ELS{}
$\bar{\mathcal{B}}$ with dimension greater than $n-u$. Then there
exists an \ELS{} $\mathcal{B}$ with dimension $< u$ such that ${\bf
x} \in \mathcal{B}$, which contradicts Lemma~\ref{lemma:x_in_A}.
\end{proof}

The decomposition over two complementary ELS's induces a mapping
from $\mbox{GF}(q^m)^n$ to $\mbox{GF}(q^m)^v$.

\begin{definition}
Let $\mathcal{V} \in E_v(q^m,n)$ be the row span of ${\bf B}$ with
an elementary complement $\bar{\mathcal{V}}$. For any ${\bf x} \in
\mathrm{GF}(q^m)^n$, we define $r_\mathcal{V}({\bf x})
=(r_0,\ldots,r_{v-1}) \in \mathrm{GF}(q^m)^v$ to be
$r_\mathcal{V}({\bf x}) = {\bf x}_\mathcal{V} {\bf B}^{-R}$, where
${\bf B}^{-R}$ is the right inverse of ${\bf B}$.
\end{definition}

We remark that the $r_\mathcal{V}$ function is linear and since
${\bf B}^{-R}$ has full rank, we have $\rk(r_\mathcal{V}({\bf x})) =
\rk({\bf x}_\mathcal{V})$ for all ${\bf x}$.

\begin{definition}
For $\mathcal{V} \in E_v(q^m,n)$ the row span of ${\bf B}$, let
$\bar{\mathcal{V}}$ be an elementary complement of $\mathcal{V}$
which is the row span of $\bar{\bf B}$. For any ${\bf x} \in
\mathrm{GF}(q^m)^n$, we define
$s_{\mathcal{V},\bar{\mathcal{V}}}({\bf x}) =(r_\mathcal{V}({\bf
x}),r_{\bar{\mathcal{V}}}({\bf x})) \in \mathrm{GF}(q^m)^n$.
\end{definition}

\begin{lemma}\label{lemma:s_V,W}
For all ${\bf x} \in \mathrm{GF}(q^m)^n$,
$\rk(s_{\mathcal{V},\bar{\mathcal{V}}}({\bf x})) = \rk({\bf x})$.
\end{lemma}
\begin{proof}
Note that ${\bf x} = s_{\mathcal{V},\bar{\mathcal{V}}}({\bf x})
\hat{\bf B}$ where $\hat{\bf B} = ({\bf B}, \bar{\bf B})^T$ is an $n
\times n$ matrix over $\mathrm{GF}(q)$ with full rank. Therefore
$\rk(s_{\mathcal{V},\bar{\mathcal{V}}}({\bf x})) = \rk({\bf x})$.
\end{proof}

\begin{corollary}\label{cor:rk_rv}
For all ${\bf x} \in \mathrm{GF}(q^m)^n$ and two complementary ELS's
$\mathcal{V}$ and $\bar{\mathcal{V}}$, $0\leq \rk({\bf
x}_\mathcal{V})\leq \rk({\bf x})$ and $\rk({\bf x})\leq \rk({\bf
x}_\mathcal{V})+\rk({\bf x}_{\bar{\mathcal{V}}})$.
\end{corollary}
It can be easily shown that the second inequality in
Corollary~\ref{cor:rk_rv} can be strict in some cases. For example,
consider ${\bf x} = \left(1, \, 1\right)\in \mathrm{GF}(q^2)^2$. For
appropriate ELS's $\mathcal{V}$ and $\bar{\mathcal{V}}$, ${\bf
x}_\mathcal{V}=\left(1, \, 0\right)$ and ${\bf
x}_{\bar{\mathcal{V}}}=\left(0, \, 1\right)$. Clearly, $\rk({\bf
x})<\rk({\bf x}_\mathcal{V})+\rk({\bf x}_{\bar{\mathcal{V}}})$.
However, when $\mathcal{V}$ and $\bar{\mathcal{V}}$ are two
complementary sets of coordinates, the Hamming weight of any vector
${\bf x} \in \mathrm{GF}(q^m)^n$ is the sum of the Hamming weights
of the projections of ${\bf x}$ on $\mathcal{V}$ and
$\bar{\mathcal{V}}$. Therefore, Corollary~\ref{cor:rk_rv}
illustrates the difference between ELS's and sets of coordinates.

\subsection{Properties of MRD codes}\label{sec:properties_MRD}
We now derive some useful properties of MRD codes, which will be
instrumental in Section~\ref{sec:main_results}. These properties are
similar to those of MDS codes. Let $C$ be an MRD code over
$\mathrm{GF}(q^m)$ with length $n$ ($n \leq m$), cardinality
$q^{mk}$, redundancy $r=n-k$, and minimum rank distance
$d_{\mbox{\tiny R}} = n-k+1$. We emphasize that $C$ may be {\em
linear or nonlinear}, which is necessary for our derivation in
Section~\ref{sec:main_results}. First, we derive the basic
combinatorial property of MRD codes.
\begin{lemma}[Basic combinatorial property]
\label{lemma:combinatorial_prop} For any $\mathcal{K} \in
E_k(q^m,n)$ and its elementary complement $\bar{\mathcal{K}}$ and
any vector ${\bf k} \in \mathcal{K}$, there exists a unique codeword
${\bf c} \in C$ such that ${\bf c}_\mathcal{K} = {\bf k}$.
\end{lemma}
\begin{proof}
Suppose there exist ${\bf c}, {\bf d} \in C, {\bf c} \neq {\bf d}$
such that ${\bf c}_\mathcal{K} = {\bf d}_\mathcal{K}$. Then ${\bf c}
- {\bf d} \in \bar{\mathcal{K}}$, and $0<\rk({\bf c} - {\bf d}) \leq
n-k$ by Proposition~\ref{prop:rk=dim}, which contradicts the fact
that $C$ is MRD. Then all the codewords lead to different
projections on $\mathcal{K}$. Since $|C| =|\mathcal{K}| = q^{mk}$,
for any ${\bf k} \in \mathcal{K}$ there exists a unique ${\bf c}$
such that ${\bf c}_\mathcal{K} = {\bf k}$.
\end{proof}

Lemma~\ref{lemma:combinatorial_prop} allows us to bound the rank
distribution of MRD codes.

\begin{lemma}[Bound on the rank distribution]\label{lemma:bound_Au}
Let $A_u$ be the number of codewords in $C$ with rank $u$. Then, for
$u \geq d_{\mbox{\tiny R}}$,
\begin{equation}\label{eq:bound_Au}
    A_u \leq {n \brack u}A(m,u-r).
\end{equation}
\end{lemma}
\begin{proof}
By Lemma~\ref{lemma:vanish}, any codeword ${\bf c}$ with rank $u
\geq d_{\mbox{\tiny R}}$ vanishes on an \ELS{} with dimension
$v=n-u$. Thus (\ref{eq:bound_Au}) can be established by first
determining the number of codewords vanishing on a given \ELS{} of
dimension $v$, and then multiplying by the number of such \ELS{}'s,
${n \brack v}$=${n \brack u}$. For $\mathcal{V} \in E_v(q^m,n)$,
$\mathcal{V}$ is properly contained in an \ELS{} $\mathcal{K}$ with
dimension $k$ since $v \leq k-1$. By
Lemma~\ref{lemma:combinatorial_prop}, ${\bf c}$ is completely
determined by ${\bf c}_\mathcal{K}$. Given an elementary basis of
$\mathcal{K}$ having $v$ elements that span $\mathcal{V}$, it
suffices to determine $r_\mathcal{K}({\bf c})$. However, ${\bf
c}_\mathcal{K}$ vanishes on $\mathcal{V}$, hence $v$ of the
coordinates of $r_\mathcal{K}({\bf c})$ must be zero. By
Lemma~\ref{lemma:vanish}, the other $k-v$ coordinates must be
nonzero, and since $\rk({\bf c}_\mathcal{K}) = n-v$, these
coordinates must be linearly independent. Hence, a codeword that
vanishes on $\mathcal{V}$ is completely determined by $k-v$
arbitrary linearly independent coordinates. There are at most
$A(m,k-v) = A(m,u-r)$ choices for these coordinates, and hence at
most $A(m,u-r)$ codewords that vanish on $\mathcal{V}$.
\end{proof}

Note that the exact formula for the rank distribution of {\em
linear} MRD codes was derived independently in \cite{delsarte_jct78}
and \cite{gabidulin_pit0185}. Thus, tighter bounds on $A_u$ can be
derived for linear codes. However, our derivation of the DEP of MRD
codes in Section~\ref{sec:main_results} requires bounds on $A_u$ for
both linear and nonlinear MRD codes. Therefore, the exact rank
distribution of linear MRD codes cannot be used, and the bound in
(\ref{eq:bound_Au}) should be used instead.

\begin{definition}[Restriction of a code]
For $\mathcal{V} \in E_v(q^m,n)$ $(k \leq v \leq n)$ with elementary
basis $B$ and its elementary complement $\bar{\mathcal{V}}$,
$C_\mathcal{V} =\{ r_\mathcal{V}({\bf c}) | {\bf c} \in C\}$ is
called the restriction of $C$ to $\mathcal{V}$.
\end{definition}

It is well known that a punctured MDS code is an MDS code
\cite{macwilliams_book77}. We now show that the restriction of an
MRD code to an \ELS{} is also MRD.

\begin{lemma}[Restriction of an MRD code]\label{lemma:restriction_MRD}
For all \ELS{} $\mathcal{V}$ with dimension $v$ ($k \leq  v \leq
n$), $C_\mathcal{V}$ is an MRD code with length $v$, cardinality
$q^{mk}$, and minimum rank distance $d_{\mbox{\tiny R}} = v-k+1$
over $\mathrm{GF}(q^m)$.
\end{lemma}

\begin{proof}
For ${\bf c} \neq {\bf d}\in C$, consider ${\bf x} = {\bf c} - {\bf
d}$. By the property of $r_\mathcal{V}$ function, we have
$\rk(r_\mathcal{V}({\bf c})-r_\mathcal{V}({\bf d})) =
\rk(r_\mathcal{V}({\bf c}-{\bf d})) =\rk({\bf x}_\mathcal{V}) \geq
\rk({\bf x}) - \rk({\bf x}_{\bar{\mathcal{V}}})\geq n-k+1 - (n-v) =
v-k+1$. Therefore, $C_\mathcal{V}$ is a code over $\mathrm{GF}(q^m)$
with length $v$, cardinality $q^{mk}$, and minimum rank distance
$\geq v-k+1$. The Singleton bound on $C_\mathcal{V}$ completes the
proof.
\end{proof}

\section{Performance of MRD Codes}\label{sec:main_results}
We evaluate the error performance of MRD codes using a bounded rank
distance decoder. We assume that the errors are additive and that
all errors with the same rank are equiprobable. A bounded rank
distance decoder produces a codeword within rank distance $t
=\lfloor (\dr-1)/2 \rfloor$ of the received word if it can find one,
and declares a decoder failure if it cannot. In the following, we
first derive bounds on the DEP assuming the error has rank $u$. In
the end, we derive a bound on the DEP that does \emph{not} depend on
$u$. We denote the probabilities of decoder error and failure for
the bounded rank distance decoder --- for error correction
capability $t$ and an error of rank $u$ --- as $P_E(t; u)$ and
$P_F(t; u)$ respectively. Clearly, $P_F(t; u)=P_E(t; u)=0$ for $u
\leq t$ and $P_E(t; u)=0$ and $P_F(t; u)=1$ for $t < u <
d_{\mbox{\tiny R}}-t$, which occurs only if $d_{\mbox{\tiny
R}}=2t+2$. Thus we investigate the case where $u \geq d_{\mbox{\tiny
R}}-t$ and $P_E(t; u)$ characterizes the performance of the code, as
$P_E(t; u) + P_F(t; u) = 1$.

Since our derivation below is transparent to the transmitted
codeword, we assume without loss of generality that the all-zero
vector is a codeword and is transmitted. Thus, the received word can
be any vector with rank $u$ with equal probability. We call a vector
decodable if it lies within rank distance $t$ of some codeword. If
$D_u$ denotes the number of decodable vectors of rank $u$, then for
$u \geq d_{\mbox{\tiny R}}-t$ we have
\begin{equation}\label{eq:PE_Du_Nu}
    P_E(t;u) = \frac{D_u}{N_u}=\frac{D_u}{{n \brack u}A(m,u)}.
\end{equation}
Hence the main challenge is to derive upper bounds on $D_u$. We
consider two cases, $u \geq d_{\mbox{\tiny R}}$ and $d_{\mbox{\tiny
R}}-t \leq u < d_{\mbox{\tiny R}}$, separately.

\begin{proposition}\label{prop:Du_u_geq_d}
For $u \geq d_{\mbox{\tiny R}}$, then $D_u \leq {n \brack
u}A(m,u-r)V_t$, where $V_t=\sum_{i=0}^t N_i$ is the volume of a ball
of rank radius $t$.
\end{proposition}
\begin{proof}
Any decodable vector can be uniquely written as ${\bf c} + {\bf e}$,
where ${\bf c} \in C$ and $\rk({\bf e}) \leq t$. For a fixed ${\bf
e}$, $C + {\bf e}$ is an MRD code, which
satisfies~(\ref{eq:bound_Au}). Therefore, the number of decodable
words of rank $u$ is at most ${n \brack u}A(m,u-r)$, by
Lemma~\ref{lemma:bound_Au}, multiplied by the number of error
vectors, $V_t$.
\end{proof}

\begin{lemma}\label{lemma:choices_z}
Suppose ${\bf y}=(y_0,\ldots,y_{v-1}) \in \mathrm{GF}(q^m)^v$ has
rank $w$. Then there exist ${u \brack s-w} A(m-w,s-w) q^{wu}$
vectors ${\bf z} = (z_0,\ldots,z_{u-1}) \in \mathrm{GF}(q^m)^u$ such
that ${\bf x}= ({\bf y}, {\bf z}) \in \mathrm{GF}(q^m)^{u+v}$ has
rank $s$.
\end{lemma}

\begin{proof}
Let $\mathfrak{T}$ be an ($m-w$)-dimensional subspace of
$\mathrm{GF}(q^m)$ such that $\mathfrak{T} \oplus \mathfrak{S}({\bf
y}) = \mathrm{GF}(q^m)$. We can thus express ${\bf z}$ as ${\bf z} =
{\bf a} + {\bf b}$, where $a_i \in \mathfrak{T}$ and $b_i \in
\mathfrak{S}({\bf y})$ for all $i$. Since $\rk({\bf x}) = \rk({\bf
y}) + \rk({\bf a})$, we have $\rk({\bf a}) = s-w$, and hence there
are ${u \brack s-w} A(m-w,s-w)$ possible choices for ${\bf a}$.
Also, there are $q^{wu}$ choices for the vector ${\bf b}$.
\end{proof}

We also obtain a bound similar to the one in
Proposition~\ref{prop:Du_u_geq_d} for $d_{\mbox{\tiny R}}-t \leq u <
d_{\mbox{\tiny R}}$.

\begin{proposition}\label{cor:bound_Du}
For $d_{\mbox{\tiny R}}- t\leq u < d_{\mbox{\tiny R}}$, then $D_u <
\frac{q^2}{q^2-1}{n \brack u}(q^m-1)^{u-r}V_t$.
\end{proposition}
\begin{proof}
Recall that a decodable vector of rank $u$ can be expressed as ${\bf
c} + {\bf e}$, where ${\bf c} \in C$ and $\rk({\bf e}) \leq t$. This
decodable vector vanishes on an \ELS{} $\mathcal{V}$ with dimension
$v = n-u$ by Lemma~\ref{lemma:vanish}. We have $w \df
\rk(r_\mathcal{V}({\bf c})) \leq t$ by Corollary~\ref{cor:rk_rv}.
$C_\mathcal{V}$ is an MRD code by Lemma~\ref{lemma:restriction_MRD},
hence $w \geq d_{\mbox{\tiny R}}-u$. A codeword ${\bf c} \in C$ is
completely determined by ${\bf c}_\mathcal{V}$ by
Lemma~\ref{lemma:restriction_MRD}. Denoting $r'=r-u$, the number of
codewords in $C_\mathcal{V}$ with rank $w$ is at most ${v \brack
w}A(m,w-r')$ by Lemma~\ref{lemma:bound_Au}. For each codeword ${\bf
c}$ such that $\rk(r_\mathcal{V}({\bf c})) = w$, we count the number
of error vectors ${\bf e}$ such that $r_\mathcal{V}({\bf c}) +
r_\mathcal{V}({\bf e}) = {\bf 0}$. Suppose that ${\bf e}$ has rank
$s$ $(w \leq s \leq t)$, then
$s_{\mathcal{V},\bar{\mathcal{V}}}({\bf e}) = (-r_\mathcal{V}({\bf
c}),r_{\bar{\mathcal{V}}}({\bf e}))$ has rank $s$ by
Lemma~\ref{lemma:s_V,W}. By Lemma~\ref{lemma:choices_z}, there are
at most ${u \brack s-w} A(m-w,s-w) q^{wu}$ choices for
$r_{\bar{\mathcal{V}}}({\bf e})$, and hence as many choices for
${\bf e}$.

The total number $D_\mathcal{V}$ of decodable vectors vanishing on
$\mathcal{V}$ is then at most
\begin{eqnarray}
    \nonumber
    D_\mathcal{V} &\leq& \sum_{w=d_{\mbox{\tiny R}}-u}^t {v \brack w}
    A(m,w-r')\\
    \label{eq:DV_u<d}
    &&\cdot \sum_{s=w}^t {u \brack s-w} A(m-w,s-w) q^{wu}.
\end{eqnarray}
We have $A(m,w-r+u) \leq (q^m-1)^{w-r+u}$ and $q^{wu}A(m-w,s-w) \leq
q^{w(u-s+w)} A(m,s-w)$. Equation~(\ref{eq:DV_u<d}) implies
\begin{eqnarray}
    \nonumber
    D_\mathcal{V} &\leq& (q^m-1)^{u-r} \sum_{s=w}^t \sum_{w=d_{\mbox{\tiny R}}-u}^s
    {v \brack w} {u \brack s-w}\\
    \nonumber
    &&\cdot q^{w(u-s+w)} A(m,s-w)(q^m-1)^w\\
    \nonumber
    &<& (q^m-1)^{u-r}\sum_{s=d_{\mbox{\tiny R}}-u}^t q^{ms}\\
    \nonumber
    &&\cdot \sum_{w=d_{\mbox{\tiny R}}-u}^s
    {v \brack w} {u \brack s-w} q^{w(u-s+w)}.
\end{eqnarray}
Using \cite[p. 225]{andrews_book76}: $\sum_{w=0}^s {v \brack w} {u
\brack s-w} q^{w(u-s+w)} = {v+u \brack s}$, we obtain $D_\mathcal{V}
< (q^m-1)^{u-r}\sum_{s=d_{\mbox{\tiny R}}-u}^t q^{ms} {n \brack s}$.
By Lemma~\ref{lemma:bound_Amu}, we find that $D_\mathcal{V} <
(q^m-1)^{u-r}\sum_{s=d_{\mbox{\tiny R}}-u}^t \frac{q^2}{q^2-1}
A(m,s) {n \brack s} < \frac{q^2}{q^2-1} (q^m-1)^{u-r} V_t$.

The result follows by multiplying the bound on $D_\mathcal{V}$ by
${n \brack v}$, the number of \ELS{}'s of dimension $v$.
\end{proof}

Finally, we can derive our bounds on the DEP.

\begin{proposition}\label{prop:bound_PE1}
For $d_{\mbox{\tiny R}}-t \leq u < d_{\mbox{\tiny R}}$, the DEP
satisfies
\begin{equation}\label{eq:bound_PE11}
    P_E(t;u) < \frac{q^2}{q^2-1} \frac{(q^m-1)^{u-r}}{A(m,u)}V_t.
\end{equation}
For $u \geq d_{\mbox{\tiny R}}$, the DEP satisfies
\begin{equation}\label{eq:bound_PE12}
    P_E(t;u) < \frac{A(m,u-r)}{A(m,u)}V_t.
\end{equation}
\end{proposition}
\begin{proof}
The bound in (\ref{eq:bound_PE11}) follows directly
from~(\ref{eq:PE_Du_Nu}) and Proposition~\ref{cor:bound_Du}, while
the bound in (\ref{eq:bound_PE12}) follows directly
from~(\ref{eq:PE_Du_Nu}) and Proposition~\ref{prop:Du_u_geq_d}.
\end{proof}

The result may be weakened in order to find a bound on the DEP in
exponential form which depends on $t$ only. In order to obtain this
bound, we need a bound on $V_t$ first.
\begin{lemma}\label{lemma:bound_Vt}
For $0 \leq t \leq \min\{n,m\}$, $V_t \leq {n \brack t} q^{mt} <
K_q^{-1} q^{t(m+n-t)}$.
\end{lemma}
\begin{proof}
Without loss of generality, assume the ball is centered at zero.
From Lemma~\ref{lemma:x_in_A}, every vector ${\bf x}$ in the ball
belongs to some \ELS{} $\mathcal{V}$ with dimension $t$. Since
$|\mathcal{V}| = q^{mt}$ and $|E_t(q^m,n)| = {n \brack t}$, it
follows that $V_t \leq {n \brack t}q^{mt}$. By
Corollary~\ref{cor:bound_gaussian}, we have ${n \brack t} < K_q^{-1}
q^{t(n-t)}$, and hence $V_t < K_q^{-1} q^{t(m+n-t)}$.
\end{proof}

\begin{proposition}\label{prop:bound_PE2}
For $u \geq d_{\mbox{\tiny R}}-t$, the DEP satisfies
\begin{equation}\label{eq:bound_PE3}
    P_E(t;u) < \frac{q^{-t^2}}{K_q^{2}}.
\end{equation}
\end{proposition}
\begin{proof}
First suppose that $u \geq d_{\mbox{\tiny R}}$. Applying $A(m,u) >
K_q q^{mu}$ in Lemma~\ref{lemma:bound_Amu} and
Lemma~\ref{lemma:bound_Vt} to (\ref{eq:bound_PE12}), we obtain
$P_E(t;u) < K_q^{-2} q^{-mr+t(m+n-t)}$. Since $n \leq m$ and $2t
\leq r$, it follows that $P_E(t;u) < K_q^{-2} q^{-t^2}$. For
$d_{\mbox{\tiny R}}-t \leq u < d_{\mbox{\tiny R}}$, applying $A(m,u)
> \frac{q}{q-1} K_q q^{mu}$ in Lemma~\ref{lemma:bound_Amu}
and Lemma~\ref{lemma:bound_Vt} to (\ref{eq:bound_PE11}), we obtain
$P_E(t;u) < \frac{q^2(q-1)}{q(q^2-1)} K_q^{-2} q^{-mr+t(m+n-t)} <
K_q^{-2} q^{-t^2}$.
\end{proof}

Based on the proof above, it is clear that the bound in
Proposition~\ref{prop:bound_PE2} is less tight than those in
Proposition~\ref{prop:bound_PE1}. However, the bound in
Proposition~\ref{prop:bound_PE2} does not depend on the rank of the
error at all. This implies that the bound applies to any error
vector provided the errors with the same rank are equiprobable.
Based on conditional probability, we can easily establish
\begin{corollary}\label{cor:anyerror}
For an MRD code with $d_{\mbox{\tiny R}}=2t+1$ and any additive
error such that the errors with the same rank are equiprobable, the
DEP of a bounded rank distance decoder satisfies $P_E(t) < K_q^{-2}
q^{-t^2}.$
\end{corollary}

\section{Acknowledgment}
The authors are grateful to the anonymous reviewers and the
associate editor Dr.~Ludo Tolhuizen for their constructive comments,
which improve both the results and the presentation of the paper. We
wish to thank one of the reviewers for suggesting the present proofs
for Proposition~\ref{prop:complementary_ELS} and
Lemma~\ref{lemma:bound_Au}, which are more concise than our original
proofs. Due to another reviewer's comments, the precise expression
in Lemma~\ref{lemma:choices_z} replaced an upper bound in our
original submission.


%
%
%

\end{document}